\documentclass[twocolumn,superscriptaddress,aps,prl]{revtex4-2}

\usepackage{graphicx}
\usepackage{dcolumn}
\usepackage{bm}
\usepackage[breaklinks,hidelinks]{hyperref}
\begin{document}

\preprint{AIP/123-QED}

\title{Sensing Strain-induced Symmetry Breaking by Reflectance Anisotropy Spectroscopy}

\author{M. Volpi*}
\email{marco.volpi@mat.ethz.ch}
\affiliation{ETH Zurich, Department of Materials, Laboratory for Nanometallurgy, Vladimir-Prelog-Weg 5, Zurich, 8093, Switzerland
}%
\author{S. Beck*}%
\affiliation{ETH Zurich, Department of Materials, Materials Theory, Wolfgang-Pauli-Str. 27, Zurich, 8093, Switzerland
}%
\affiliation{Center for Computational Quantum Physics, Flatiron Institute, 162 5th Avenue, New York, NY 10010, USA
}%
\author{A. Hampel}%
\affiliation{Center for Computational Quantum Physics, Flatiron Institute, 162 5th Avenue, New York, NY 10010, USA
}%
\author{H. Galinski}%
\affiliation{ETH Zurich, Department of Materials, Laboratory for Nanometallurgy, Vladimir-Prelog-Weg 5, Zurich, 8093, Switzerland
}%
\author{A. Sologubenko}%
\affiliation{ETH Zurich, Department of Materials, Laboratory for Nanometallurgy, Vladimir-Prelog-Weg 5, Zurich, 8093, Switzerland
}%
\affiliation{ETH Zurich, Scientific Center for Optical and Electron Microscopy, ScopeM, Vladimir-Prelog-Weg 5, Zürich, 8093, Switzerland 
}%
\author{R. Spolenak}%
\affiliation{ETH Zurich, Department of Materials, Laboratory for Nanometallurgy, Vladimir-Prelog-Weg 5, Zurich, 8093, Switzerland
}%
\date{\today}
\begin{abstract}
Intentional breaking of the lattice symmetry in solids is a key concept to alter the properties of materials by modifying their electronic band structure. 
However, the correlation of strain-induced effects and breaking of the lattice symmetry is often indirect, resorting to vibrational spectroscopic techniques such as Raman scattering.
Here, we demonstrate that reflectance anisotropy spectroscopy (RAS), which directly depends on the complex dielectric function, enables the direct observation of electronic band structure modulation.
Studying the strain-induced symmetry breaking in copper, we show how uniaxial strain lifts the degeneracy of states in the proximity of the both L and X symmetry points, thus altering the matrix element for interband optical transitions, directly observable in RAS. We corroborate our experimental results by analysing the strain-induced changes in the electronic structure based on ab-initio density functional theory calculations. 
The versatility to study breaking of the lattice symmetry by simple reflectance measurements opens up the possibility to gain a direct insight on the band-structure of other strain-engineered materials, such as graphene and two-dimensional (2D) transition metal dichalcogenides (TMDCs).

\end{abstract}

\maketitle

Strain engineering, i.e. the controlled deformation of solid state materials, is an elegant and versatile method to unlock new material properties. Thus, strain-induced symmetry breaking has gained considerable attention in condensed matter physics, enabling the observation of the anomalous Hall effect (AHE) in magnetic thin films~\cite{Kim2020StrainFilms}, breaking the absorption limit of silicon \cite{Katiyar2020BreakingEngineering} and band-gap engineering of halide perovskites \cite{Chen2020StrainPerovskites}. Moreover, the concept proves most helpful to tune the band structure of 2D materials \cite{K.S.NovoselovA.K.GeimS.V.MorozovD.JiangY.ZhangS.V.Dubonos2016ElectricFilms, Bai2020ExcitonsHeterojunctions,Hsu2020NanoscaleGraphene}, where strain engineering combined with the outstanding mechanical flexibility of 2D materials enables tailorable enhanced electronic and optical properties.

\begin{figure*}[t!]
    \centering
    \includegraphics[width=\textwidth]{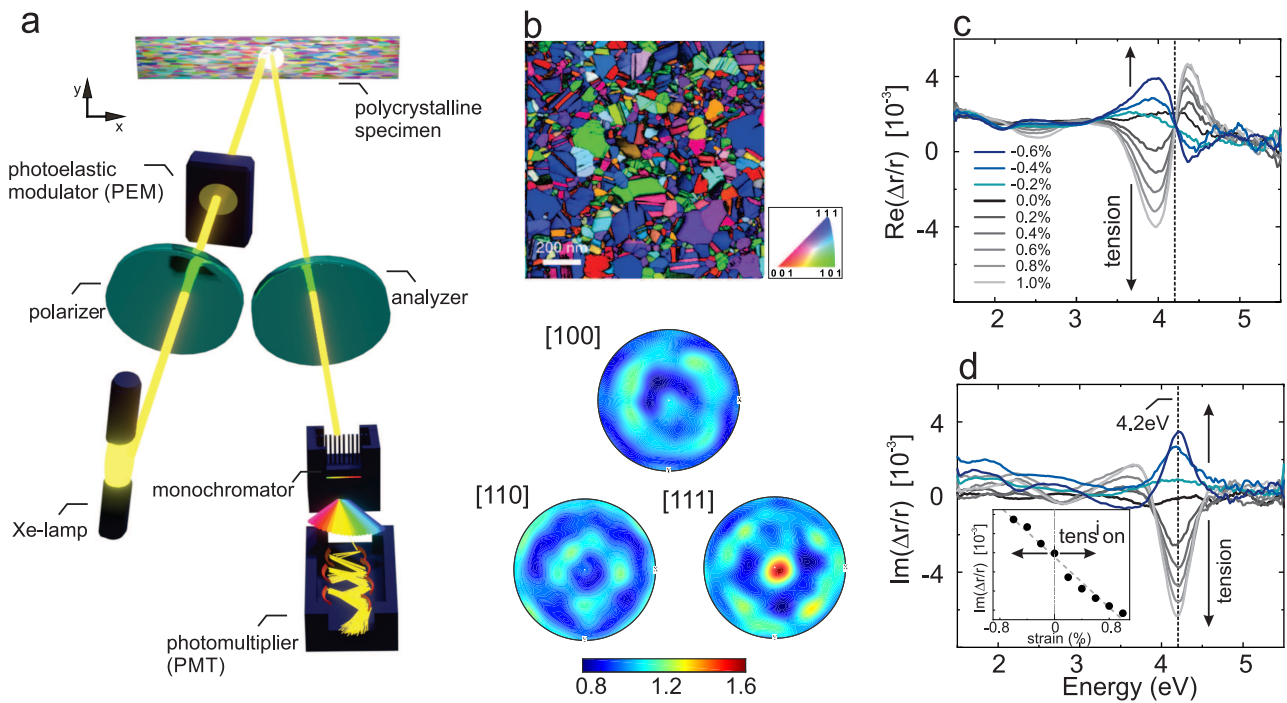}
    \caption{RAS setup and spectra. RAS is composed of a Xe-lamp, a collimator and a polarizer along the incident beam direction that polarize the light beam at 45° with respect to the \textit{x} and \textit{y} specimen axes. Along the reflected beam direction, it features a photoelastic modulator and an analyser to modulate the phase of the light, which is eventually measured at different wavelengths by means of a monochromator and a photomultiplier. The annealed Cu polycrystalline specimen is reported in an EBSD fashion. b) reports the sample’s orientation map, where colours represent out-of-plane crystallographic orientations as in the legend, and the relative pole figures along the 3 main crystallographic axes. c) and d) are the real and imaginary part, respectively, of the RAS signal for a set of uniaxial tensile and compressive strains. The absence of signal at 0 \% strain is due to an overall in-plane optical isotropy. The appearance of the elastic feature for uniaxial strains higher than 0 \% – indicated by the black dashed line – is due to symmetry break. The inset in d) shows the linear relation between the $\Delta{r}/{r} $ at the elastic feature energy (~4.2 eV) with the strain, ensuring that yield has not been reached.}
    \label{fig:1}
\end{figure*}

For instance, single layer MoS$_2$ subjected to uniform uniaxial and biaxial tensile strain shows its bandgap linearly decreasing with increasing tensile strain, first undergoing a direct-to-indirect bandgap transition and then a semiconductor-to-metal transition \cite{Scalise2012Strain-inducedMoS2,Ghorbani-Asl2013Strain-dependentDichalcogenides}. Also, strain abruptly modifies the electronic band structure of graphene, leading to the appearance of an optical band gap \cite{Guinea2010EnergyEngineering} accompanied by a change in the optical conductivity \cite{Pellegrino2010StrainGraphene}.\\
\\
However, the measurement of the strain state and the correlation of the reported strain-induced effects with the breaking of the lattice symmetry is often indirect. Commonly, Raman scattering is employed, which allows for an indirect characterization of strain due to shifting vibrational modes \cite{Dai2019StrainInterface}. A determination of the single strain components is not straightforward and requires additional information about the stress state.\\
In this Letter, we demonstrate that reflectance anisotropy spectroscopy (RAS), thanks to its direct correlation to the dielectric function, enables the direct observations of electronic band structure modulation due to strain-induced symmetry breaking. We employ density functional theory (DFT) calculations to explain and characterize the emergence of the RAS signal as  result of the symmetry breaking. RAS, also known as reflectance difference spectroscopy (RDS) \cite{Weightman2005ReflectionSpectroscopy,Ronnow1999DeterminationSpectroscopy,Martin2001ReflectionSurfaces}, is a non-destructive phase modulating technique capable of detecting materials optical anisotropy (Fig.~\ref{fig:1}a). RAS measures the complex reflection anisotropy $\Delta r/r$, defined as the normalized difference of complex frequency dependent Fresnel reflection amplitudes $r_x$ and $r_y$ of two orthogonal directions $x$ and $y$:
\begin{equation}
 \frac{\Delta r}{r} = 
  \frac{r_{x}-r_{y}}{r_{x}+r_{y}} \quad .
  \label{eq:1}
\end{equation}

RAS, developed by Aspnes et al. \cite{Aspnes1985AnisotropiesSemiconductors}, is a versatile method to study surface reconstruction \cite{Aspnes1988ApplicationAlAs}, dimer orientations \cite{Power1998SensitivitySi001}, epitaxial growth \cite{Aspnes1988ApplicationAlAs}, piezo-optical properties \cite{Ronnow1999DeterminationSpectroscopy}, and thin film growth \cite{Martin2001ReflectionSurfaces}. To date, it also extends to surface physics and chemistry, enabling to resolve phenomena such as surface molecular absorption \cite{Martin2001ReflectionSurfaces}.\\
Intuitively, cubic crystallographic orientations like [1 1 1] and [1 0 0] do not contribute to $\Delta r/r $ as $ r_{x}=r_{y} $ nor do polycrystalline specimens with random grain orientations whose in-plane anisotropy averages out \cite{Weightman2005ReflectionSpectroscopy}. Therefore, even though largely employed in industry, polycrystalline specimens have been disregarded for long. 
The RA signal is often linked to band structure modulation [14–17] and first ab-initio DFT simulations confirm this \cite{Harl2007AbSurfaces}, focusing on surface states in single-crystalline materials. However, a comprehensive study linking the strain-induced reflection anisotropy spectra to symmetry breaking in the band structure and a general understanding of its physical origin is still lacking.\\ 
\\
Here, we probe the strain-induced symmetry breaking using the specific example of Cu thin films. 200 nm thick polycrystalline Cu films were deposited using magnetron assisted physical vapour deposition (PVD) on strain-free polyimide substrate (Kapton E\textsuperscript{\textregistered} Du Pont) (see also supplementary material \cite{SM}). The choice of pure Cu, with a plasma frequency at ~7.2 eV, excludes absorptions caused by non-metal-like behavior within the explored energy range (1.5-5.5 eV) \cite{H.1962OpticalCu}, and facilitates an undisturbed study of interband transitions. The same accounts for free electron contributions (Drude model) which only play a role for energies below 1.5 eV and hence can be neglected. Moreover, these films are easy to produce and simpler to strain than e.g. 2D materials \cite{Dai2019StrainInterface}; also, they guarantee a high surface quality and flatness that do not cause topography related contributions to the RA signal [Fig. S1]. 
Prior to analysis, thin films were annealed in vacuum at 300°C for 24 h. This thermal treatment ensures the desired (1 1 1) out-of-plane texture shown in Figure~\ref{fig:1}b, which prevents \textit{parasitic} optical contribution from inherent anisotropic surface energy states~\cite{Weightman2005ReflectionSpectroscopy}.

\begin{figure*}[t]
    \centering
    \includegraphics[width=\textwidth]{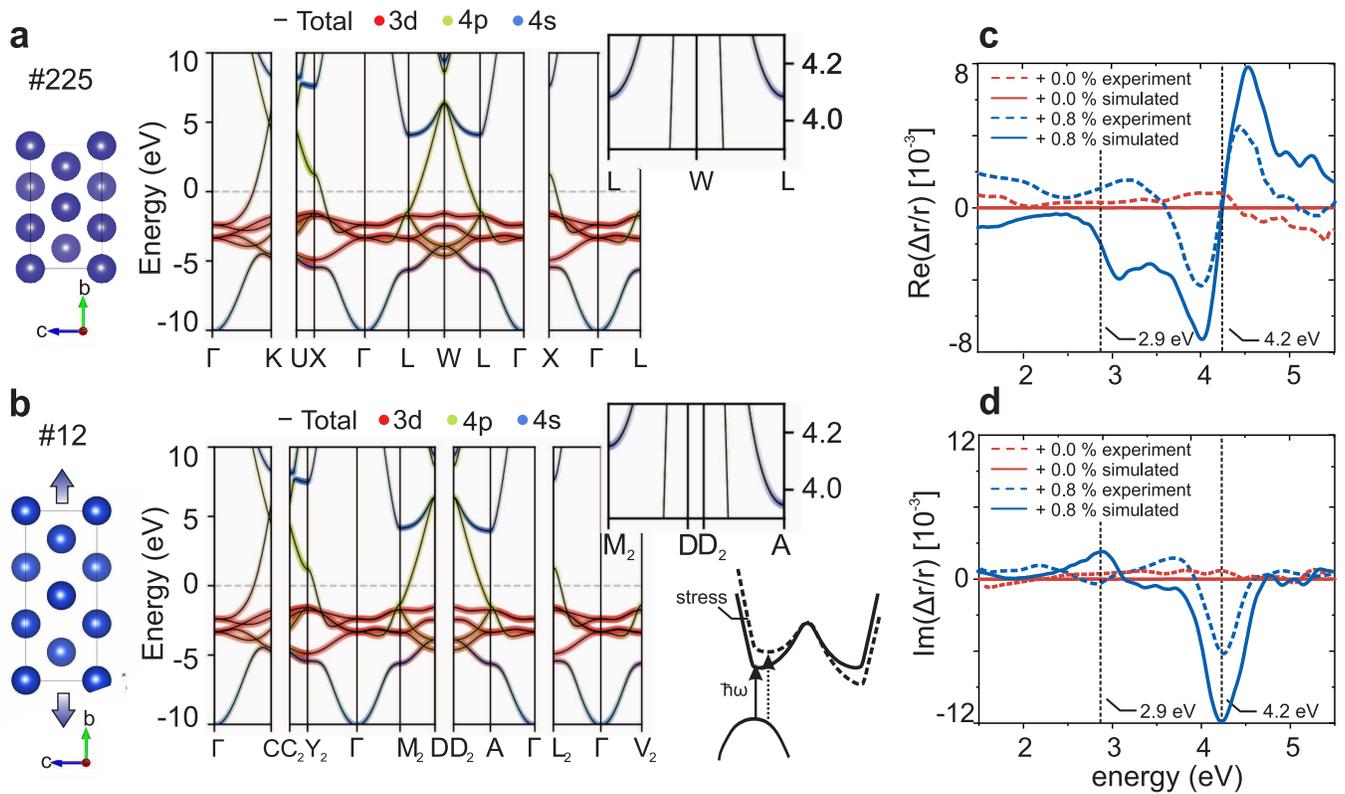}
    \caption{DFT simulations of RA signal. The figure reports the band structure obtained by DFT for (a) the unstrained  Cu unit cell (space group \#225) and (b) the elongated cell along $b$ [2 -1 -1] (space group \#12).  A schematics of a 2D representation (1 1 1) cell is represented next to the band structure. (c) and (d) display the real and imaginary components, respectively, of the simulated RA signal (solid lines) superimposed to the experimental (dashed lines) for both pristine and 0.8 \% elongated unit cell and thin film, respectively. The experimental signal is offset with respect to the $y$ axis in both (c) and (d) to facilitate the comparison with simulations. The black dashed lines highlight the features of interest at ~2.9 and ~4.2 eV. }
    \label{fig:2}
\end{figure*}

The pole figures observed in Figure~\ref{fig:1}b, indicate an in-plane random orientation. Thus, the unstrained specimens considered in this work do not carry in-plane inherent asymmetry and $ r_{x} $ and $r_{y} $ coincide. So, the numerator in Eq.~\ref{eq:1} vanishes and no RA signal is seen.
This situation changes once the films are uniaxially strained. Strain is induced on 3 x 25 mm$^2$ Cu films by using a tensile stage (Kammrath \& Weiss GmbH, Germany).
Considering the enhanced mechanical properties of thin films' fine structure defined by the Hall-Petch theory \cite{Hall1951TheResults}, we ensure elastic behavior up to 0.8 \% strain, avoiding optical contributions from dislocations. Figure~\ref{fig:1}c and \ref{fig:1}d display the real and imaginary reflectance anisotropy spectra of a Cu thin film at varying uniaxial strain within the elastic regime. The observed lineshape resembles a complex Lorentzian oscillator centered at a resonant frequency (energy). Indeed, similarly to the frequency dependence of the Lorentzian response \cite{Hummel2012}, the imaginary part Im($\Delta{r}/{r} $) shows a peak, while the real part exhibits a typical 2$^{\text{nd}}$ derivative trend, featuring a positive or negative slope depending on the strain state. Figure~\ref{fig:1}c and \ref{fig:1}d also show that the feature disappears and reverses in sign the strain state changes from tension to compression. In accordance with synchrotron measurements \cite{Wyss2017MonitoringDiffraction}, the inset in Fig.~\ref{fig:1}d  reveals that the peak of Im($\Delta{r}/{r} $) scales linearly with strain. Owing to the linear elastic response, we will refer to it as "elastic feature" throughout the text.

\begin{figure*}[t!]
    \centering
    \includegraphics[width=\textwidth]{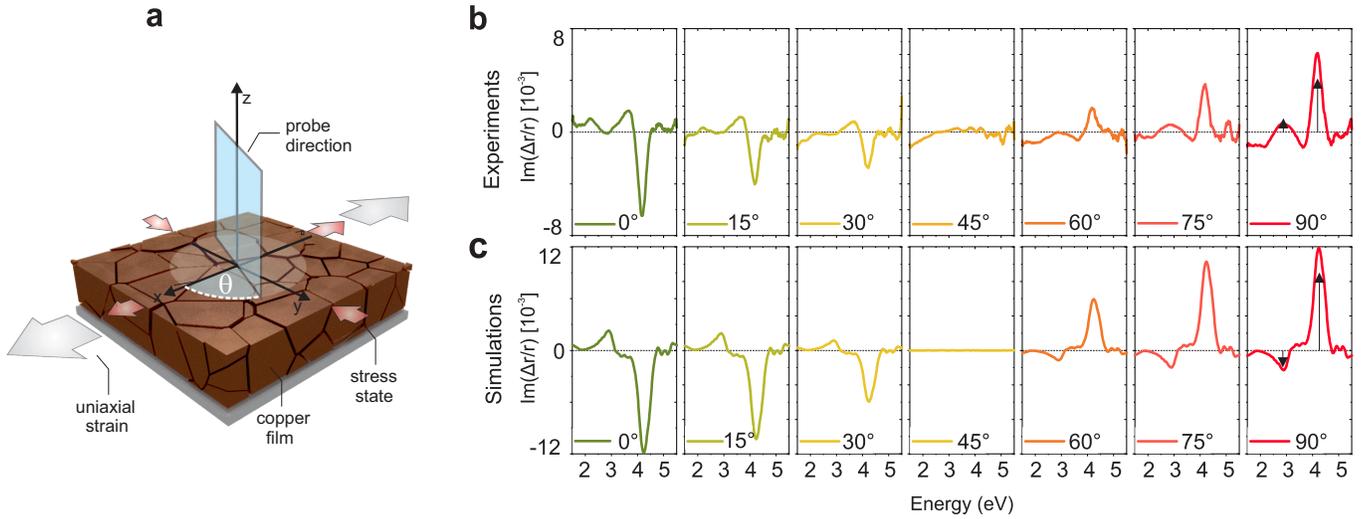}
    \caption{Azimuthal RAS signal. The schematics (a) shows the experimental approach. The polycrystalline Cu elongates along the strain direction and contracts in the perpendicular one. This induces optical anisotropy which is probed over a set of azimuthal angles $\vartheta$. The resulting imaginary component of the experimental RAS signal upon 0.8 \% strain over a set of angles is reported in (b), e.g. 0$^{\circ}$ equals $2*(r_{x} - r_{y})/(r_{x}+r_{y})$, 15° equals  $2*(r_{x+15^{\circ}} - r_{y+15^{\circ}})/(r_{x+15^{\circ}}+r_{y+15^{\circ}})$, 30° equals  $2*(r_{x+30^{\circ}} - r_{y+30^{\circ}})/(r_{x+30^{\circ}}+r_{y+30^{\circ}})$, etc. The imaginary component of the simulated RAS signal upon 0.8 \% elongation in b [1 -1 0] over the same set of  angles is reported in (c), e.g. 0$^{\circ}$ equals $2*(r_{b} - r_{c})/(r_{b}+r_{c})$, 15° equals  $2*(r_{b+15^{\circ}} - r_{c+15^{\circ}})/(r_{b+15^{\circ}}+r_{c+15^{\circ}})$, 30$^{\circ}$ equals  $2*(r_{b+30^{\circ}} - r_{c+30^{\circ}})/(r_{b+30^{\circ}}+r_{c+30^{\circ}})$, etc. The y-axis scale bars between b and c are adapted to the features' amplitude.
    }
    \label{fig:3}
\end{figure*}

To simulate the dielectric response of uniaxially strained Cu, we carry out a set of ab-initio DFT calculations using the \textsc{Quantum~ESPRESSO} software package \cite{Giannozzi2009QUANTUMMaterials}, and the exchange correlation functional according to Perdew, Burke, and Ernzerhof \cite{Perdew1996GeneralizedSimple}. We model three structures: two strained ones with the strain applied perpendicular to the [1 1 1] direction, i.e. along [2 -1 -1] and [0 1 -1] (from now on referred to as $b$ and $c$ direction, respectively), to replicate the experimental (1 1 1) texture, and the unstrained one for comparison. Details on the construction of the supercells and computational parameters are given in the supplementary information \cite{SM}. Tensile strain of a magnitude of 0.8 \% is applied in agreement to the experimental setup. It is to note that uniaxial strain perpendicular to the [1 1 1]-direction reduces the space group from \#225 of the unstrained to \#12 for the strained lattices (Figure~\ref{fig:2}a-b), which results in a breaking of symmetry visible in the bandstructure.
To obtain the reflectance anisotropy, i.e. $\Delta{r}/{r} $, at the (1 1 1) surface, the diagonal elements of the dielectric tensor $\epsilon_{\alpha \alpha}$ in Cartesian coordinates are calculated with the epsilon.x post-processing tool of \textsc{Quantum~ESPRESSO}.  We only consider interband transitions in the calculation of $\epsilon_{\alpha \beta} (\omega)$, since the Drude term is not relevant for the RAS signal in the energy range considered in this work.

The contribution of the interband transitions to the total dielectric function are directly related to the joint density of states, i.e. the sum over all vertical transitions across the Fermi level across the Brillouin zone. Consequently, its variation and in turn the RAS modulation stem from the interband transitions between occupied and unoccupied energy levels. 

To understand the origin of the calculated signal, we turn to the bandstructure in Figure~\ref{fig:2}a, which shows the projection of the Kohn-Sham states on atomic wavefunctions of bulk Cu. The path of high-symmetry points is chosen such that it allows for a simple comparison with the strained structures. With the occupied 3$d$ states in an energy range [-6,-1] eV, the main contribution of the bands crossing the Fermi level is given by the 4$p$ states, followed by the unoccupied 4$s$ states.

In the unstrained configuration, the high symmetry of the structure results in an isotropic dielectric tensor, such that the calculated RA signal (Fig.~\ref{fig:2}c-d blue solid lines) vanishes in agreement with experimental results (Fig.~\ref{fig:2}c-d red dashed lines). Figure~\ref{fig:2}b shows the band structure of the strained structure along the crystallographic $b$-axis. At first glance the overall result seems unchanged in comparison to the unstrained structure, however, due to the symmetry breaking, previously symmetry-related Kohn-Sham eigenvalues are no longer degenerate along the chosen $k$-point path. This is emphasized in the inset of both bandstructure plots, focusing on the eigenstates in an energy range of [3.9-4.3] eV along L-W-L and M$_{2}$-D$||$D$_{2}$-A in unstrained and strained structure, respectively. While the unoccupied 4$s$ states are degenerate in the former case, the uniaxial strain leads to an increase (decrease) in energy of the corresponding state at M$_{2}$ (A), leading to an energy difference of ~200 meV. 

Thus, when summing over all $k$-points in the calculation of the dielectric tensor, the transitions do not cancel out exactly. The relevant transition channel resulting in the signal are excitations from occupied 4$p$ to unoccupied 4$s$ states, which we crosschecked by performing calculations excluding these transitions in the calculation of the dielectric function (see supplementary information and Fig. S2 for details \cite{SM}). If the channel is not suppressed, it results in the observed oscillation in the real part and a dip in the imaginary part of the RA signal in an energy range of [3.9-4.6] eV, as shown in Fig.~\ref{fig:2}c-d (blue solid lines). The corresponding analysis for strain along the $c$-direction is in agreement with these results, see Fig. S4 in supplementary \cite{SM}. 
This reveals the exceptional sensitivity of the RAS to the asymmetries of the energy bands, enabling to probe band energy shifts on the meV scale.

A similar trend is also found at the high-symmetry point X in the unstrained structure, and Y$_{2}$ and L$_{2}$ in the strained one. Here, the signal results from transitions from the occupied 3$d$ to the unoccupied 4$p$ states. However, since the energy gap is smaller the resulting signal in the RA spectrum appears at smaller energies. Precisely, Figure~\ref{fig:2}c-d reports it in a range [2.7-3.2] eV with a dip at 2.9 eV. The sign is opposite in this case, as the band at Y$_{2}$ is shifted down, while at L$_{2}$ it is shifted up.

Our experimental spectra are in excellent agreement with data from literature \cite{Cole2003Stress-inducedSpectroscopy,Wyss2015ReflectanceFilms,H.1962OpticalCu,B.R.CooperH.EhrenreichH.R.Philipp1964OpticalII.}, affirming the special role of interband absoprtion due to a van Hove singularity \cite{Segall1962FermiCopper} close to the L symmetry point. Also, another transition at a joint density of states critical point in the vicinity of the X appears in the measured RA signal. 
Quite interestingly, as reported in Figure~\ref{fig:2}d, the latter exhibits sign inversion with respect to simulations.

To analyze the origin of the flipped response at ~2.9 eV, we test the robustness of the signal when the polarization direction is not aligned with the strain direction. 
Figure~\ref{fig:3} reports a comparison between experiments and DFT of the ImRAS signal for a set of azimuthal angles $\theta$ between $0^{\circ}$ and $90^{\circ}$ (schematics in Fig.~\ref{fig:3}a). 
The polycrystalline specimens elongated in the experimental reference system $x$ direction (Fig.~\ref{fig:3}a) experience optical anisotropy which is detectable at different in-plane orientations (Fig.~\ref{fig:3}b). This coincides with the trend of simulated ImRAS signal for the unit cell elongated along the $b$-direction (Fig. 3c). The sign agreement and inversion at ~4.2 eV and ~2.9 eV, respectively, is conserved over the in-plane rotation. This consistency indicates that, while the experimental band shift in the vicinity of the L symmetry point equals the simulated one, the one at the X symmetry point is opposite. This suggests that, although the same symmetries are broken in both cases, the experimental strain state differs from the calculated one. This is intuitive considering that the L symmetry point lies along the strained direction, whereas the X point does not. Thus, the uniaxial strain applied in this work directly breaks the symmetry in the vicinity of L, whereas only subsequent relaxation along one of the other directions breaks the symmetry in the vicinity of X. 

Because of this, we attribute the flip in sign of the ~2.9~eV feature to the non-ideal experimental conditions. Indeed, contrary to the simulated single crystal, polycrystalline thin films are bound to a polymeric substrate \cite{SM}, which constrains the natural transversal relaxation that would arise as a result of a Poisson ratio mismatch. This prevents the atoms distribution in specimens to reach the equilibrium, leaving residual strains on the Cu film that are not present in the calculations. Experimental confirmation using free Cu single crystals is challenging considering that the elastic limit drastically drops compared to polycrystals and a 0.8 \% strain would exceed the yield point, thus, leading to the appearance of plasticity deformation RAS contributions in the same spectral region \cite{Blackford2005RASMetals,Wyss2015ReflectanceFilms}.
Alternatively, follow-up DFT calculations without cell relaxation according to the Poisson ratio could shed light into the sign flip. However, a clear instruction of the particular state of relaxation of the experimental film remains elusive to date.
Furthermore, we note that the RAS sensitivity to the matrix element for interband optical transitions spans over 360° in-plane orientations, as emphasized in Figure~\ref{fig:3}. Therefore, although the (1 1 1) crystallographic plane exhibits a three-fold symmetry, the signal disappears at $45^{\circ}$ and it changes in sign (for constant magnitude) at a $90^{\circ}$ rotation (visible in Fig.~\ref{fig:3}b and Fig.~\ref{fig:3}c). This result not only indicates a 2-fold symmetry in the system (supplementary III \cite{SM}) but it also suggests that one can in principle determine the relative orientation of different crystallographic domains within a given material. Therefore, DFT calculations for surface \cite{Harl2007AbSurfaces} and bulk could be used to generate RA-spectra of any desired material, orientation and strain direction, paving the way for new applications.

To conclude, we have demonstrated that reflectance anisotropy spectroscopy (RAS) can serve as robust method to study symmetry breaking, as it establishes a direct link between strain state and electronic band structure in strain-engineered materials. Using the specific example of Cu we show that uniaxial strain breaks the crystal symmetry and lifts the degeneracy of the Kohn-Sham eigenvalues. Consequence of the lifted degenerancy is an anisotropic interband absorption from 4$p$ to 4$s$ states and 3$d$ and 4$p$, located in the vicinity of van Hove singularities at the X and L points, respectively. We conclude that RAS can be used to probe and resolve these strain-induced modulations of interband transitions down to the meV range, allowing for a large sensitivity to the strain direction and sign. Since RAS is versatile and can be easily adapted to study strain-engineered phenomena of other materials, this opens up possibilities for bandstructure-tuning in 2D materials \cite{Peng2020StrainApplications} and new insights into Weyl semimetals \cite{Toudert:17} and strain-controlled photoluminescence of quantum dots \cite{Shimomura2015StrainM}.

\begin{acknowledgments}
M.V. and S.B. contributed equally to this work. The authors would like to acknowledge the support from Scientific Center for Optical and Electron Microscopy (ScopeM) of ETH Zurich. This study is part of the strategic focus area advanced manufacturing project PREcision Additive Manufacturing of Precious metals Alloys (PREAMPA). The Flatiron Institute is  a division  of the Simons Foundation.
\end{acknowledgments}

\end{document}